**Cognitive mechanisms for human flocking dynamics**

**Authors:**

Seth Frey,[1, 2, 3] Robert L. Goldstone[3]

**Affiliations:**

[1] Department of Communication, University of California, Davis

[2] Neukom Institute for Computational Science, Dartmouth College

[3] Cognitive Science, Indiana University

Correspondence concerning this article, and requests for SI or data, should be addressed to Seth Frey, Dartmouth College, Moore Hall HB 6207, Hanover, NH 03755. E-mail: seth.frey@dartmouth.edu

**Abstract**

Low-level "adaptive" and higher-level "sophisticated" human reasoning processes have been proposed to play opposing roles in the emergence of unpredictable collective behaviors like crowd panics, traffic jams, and market bubbles. While adaptive processes are ubiquitous in mechanistic theories of emergent social complexity, complementary theories understand incentives, education, and other inducements to rationality as able to suppress such outcomes.

We show in a series of laboratory experiments that, rather than suppressing complex social dynamics, sophisticated reasoning processes can support them. Our experiments elicit flocking behavior in groups and show that it is driven by the human ability to recursively anticipate the reasoning of others. We identify this sophisticated flocking in three different games—the Beauty Pageant, Mod Game, and Runway Game—across which game theory predicts no formal similarity. The persistence of sophisticated flocking across unrelated game types not only speaks to the phenomenon's robustness, it also suggests that people are treating three supposedly different decision settings as conceptually similar, implicating a second sophisticated cognitive ability: human concept formation. We also find in participants' underlying reasoning that the number of recursions they perform is limited not by any individual's cognitive abilities, but by a social norm that emerges during flocking.

By implicating both recursive reasoning and concept formation in complex dynamics, we support interdisciplinary perspectives that emergent complexity is typical of even the most intelligent populations and carefully designed social systems.

**Introduction**

Cultural norms, fads and fashions, and market bubbles and busts are all unpredictable phenomena that can emerge spontaneously from disorganized human collectives.[1,2] Many mechanisms have been invoked to explain these phenomena, particularly lower-level reasoning processes such as habits, reactions, emotions, biases, reinforcement, and imitation.[3-12] Complex social dynamics are characterized by the fact that they emerge endogenously from agents' interactions and do not converge to static set of strategies. This complexity is a major impediment to the design of predictable social systems,[8,13,14] but the mechanisms that make it manageable remain poorly understood. One scientific tradition, going at least to Keynes,[15,16] holds that increasingly rational behavior can impede the emergence of complex collective behaviors by counter-acting the influence of our "animal spirits," the adaptive processes that supposedly drive social complexity.[17-22] But recent theories challenge this narrative by implicating higher-level processes in similarly complex dynamics.[23-26] If sophisticated human-unique reasoning processes can drive group outcomes spontaneously into complex dynamic regimes, then policy interventions that increase agents' motivation, knowledge, or intelligence are not sufficient to increase the simplicity or predictability of real world social dynamics.

*Sophisticated flocking in repeated games*

Theoretical findings from the nonlinear physics community have shown that simple learning mechanisms foster unexpected emergent dynamics in a wide range of economic games and models,[27-31] particularly in "large" games—those with many players, choices, and repetitions. Prominent among interesting emergent, or endogenous, dynamics is flocking, the clustering of agents in both their positions and velocities.[4,5] In repeated economic games, flocking occurs when many agents converge upon the same choices over many rounds of play.[24,29] Like other collective behaviors, flocking is usually explained in terms of low-level reasoning mechanisms,[5,32,33] and relevant studies consequently draw heavily from non-human animal collective behavior.[4,34-37]

We focus here on a collective behavior that we call "*sophisticated flocking*" that, unlike typical flocking, is driven by human higher-level reasoning mechanisms. Sophistication is clearest in iterated "what you think I think you think" reasoning,[38-41] a process that can cause people to disqualify choices that they think others will know they desire. This iterated reasoning is an expression of bounded rationality, and it is often invoked for its ability to suppress complex game dynamics associated with flocking.[17-20,42]

In many games, such as those we investigate herein, this elimination of intermediate choices makes players' actual choices appear to "hop" discontinuously through a choice space: where a basic reactive agent advances through choices sequentially by best-responding to current state, a sophisticated agent advances more quickly because it best-responds to the anticipated best-responses of others (who, in turn, are best-responding to the anticipated best response of it). In these games, a player's depth of iteration can be inferred from the number of intermediate strategies that it skips.

**Three games**

We present experimental results from three economic games, the well-known Beauty Contest and two that we developed, the Mod Game and the Runway Game.

*Beauty Contest.* In the Beauty Contest, players receive a reward for guessing the number 0–100 that is $p$ (typically $p=2/3$) of their group's mean guess.[38,39,43] The game motivates participants to try to iteratively anticipate each other's choices, and people who play it perform large hops down through the possible choices. Instructively, a researcher can infer from players' chosen numbers how many iterations they performed, and, by extension, how many iterations they expected from others: a player who responds to suspected choice 50 with 22 is, correctly or not, anticipating that others will be less sophisticated and select 33. The Beauty Contest has a unique pure-strategy Nash equilibrium, which is to say that it has a unique choice—zero—that, when chosen by everyone, leaves no one wishing they had chosen something else. While the Beauty Pageant is well known for eliciting both sophisticated hopping behaviors and low-level adaptive responses, flocking has not been suspected in the game.

*Mod Game.* In the Mod Game, $n$ players choose an integer 1–24 (or, more generally, $m$). Choice 1 dominates ("beats") choice 24 while all other choices $x$ (2–24) beat the choice $x-1$ exactly below them; players earn one point for making a choice that is exactly one above another player's, zero points otherwise, with all choices linked circularly as along the perimeter of a clock (illustrated in Figure 1.a).[24,44] This circular structure is directly reminiscent of Rock-Paper-Scissors, which the Mod Game generalizes, and with which it shares the same type of solution: a single "mixed strategy equilibrium" that players can achieve by selecting choices randomly. Frey and Goldstone find that sophistication and flocking coexist in a positive-sum version of the Mod Game.[24]

*Runway Game.* We developed the Runway Game to narrow the possible explanations for the results of the Mod Game. In both, each of multiple players earns one point for choosing a number precisely one above another's number. However, the choice space of the Runway Game is defined on the real number line, rather than on a circle of integers. Beyond having an unbounded choice space, the game is unusual because, unrepeated, it has no Nash equilibrium solution (SI). That said, it can gain one with a very small change, the addition of an "outside" choice that guarantees zero earnings. We will see that players are effectively indifferent to this change in the solvability of the basic game.

*Three game classes, three types of reasoning.* Nash equilibrium is a classic definition of the stability of game outcomes, and gives one model of rationality. But there are many kinds of game equilibria, enough that games can be classified by the solutions they have and the different methods used to find them.[45] This is important because the formal method that solves a game can not only define its type, it can also suggest the style of reasoning a player should use when playing it. For example, the mathematical method that formally solves the Beauty Contest, a model of unbounded iterated reasoning, is what first suggested to researchers that humans might themselves mentally perform iterations.[38] Just as iterated reasoning is expected in games solved this way, it is unexpected in games that are solved differently.[46-48] Specifically, neither the Mod nor the Runway game has a formal solution or structure that would predict the iterated reasoning used by players in the Beauty Contest. The formal solution to the Mod Game predicts random behavior, while the lack of a solution to the Runway Game means that there is not a game theoretic prescription for the type of reasoning that a player should employ.

**Analysis**

In the experiments reported below, all three games were repeated for many rounds, and all players received anonymized but otherwise full information about each previous round's payoff outcomes. In experiments with the Mod and Runway games, we tested both zero- and positive-sum versions, such that a player that was scored upon either did or did not lose a point for every point gained by the scoring player.

Our analyses depend on the concepts of *rate*, through which we identify sophistication, and *choice distance* and *rate distance*, with which we define flocking.

We define a player's rate in round *t* by comparing their current and previous choice. In the Mod and Runway games, rate $x'_t = x_t - x_{t-1}$. Rate greater than one is a proxy for sophistication: it reflects an iterated best-response to the anticipated iterated best-response of others. Iterated best-response is an established mechanism for the iterative elimination of adjacent series of choices that we describe as hopping.

In a game, flocking occurs when a group's members converge on shared choices and rates (illustrated in Figure 1.a). It is analogous to flocking in non-human animals, in which many animals converge to common positions and velocities. To capture convergence in both game "position" and "velocity," we measure a group's choice distance and rate distance, respectively, as the mean or median difference between group members' choices and rates within a round. We identify flocking as increasing in a group when both of these distance measures decrease significantly over time (SI). Under these definitions, sophisticated flocking is flocking with hopping (rate greater than one).

## Results

Despite the vast formal differences between these games, we robustly elicit the same *sophisticated flocking* behavior in all three.

### Mod Game

83 paid participants in 28 groups of size 2–7 played 100 rounds each of positive-sum and zero-sum versions of the Mod Game (SI. Interface in Figure S2). Groups quickly self-organized into a cyclic collective flocking behavior in which individuals "hopped" together around the strategy space (Figure 1). Players had a mean hop size of 2.2, meaning that, on average, a player who selected choice 10 in round *t* tended in round *t+1* to play choice 12, or possibly 13 (see also Figure S3). Significant decreases in groups' choice and rate distances over time ($\chi^2_1=26$; $\chi^2_1=39.2$; both $p<0.001$; Tables S2–S3, also see Figure 3) reflect a convergence upon shared "positions" and "velocities" through choice space that satisfies our definition of flocking. These distance measures were significantly larger during zero-sum conditions ($\chi^2_1=40.2$; $\chi^2_1=38.5$; both $p<0.001$; Tables S2–S3), supporting predictions that zero-sum settings will suppress complex dynamic regime in adaptive agents [28].

The rate of 2.2 is consistent with sophistication. The proximity of players to each other in both choices and rate is consistent with flocking. Sophisticated flocking is inconsistent with prominent theories of game dynamics, which propose that agents with higher-level reasoning should be less susceptible to non-equilibrium flocking in games like the Mod Game [17,18,20].

*Runway Game*

In our first experiment with this game, 59 players in 17 groups of 2–6 played two 75 round blocks of the game in zero-sum and positive-sum conditions. In a second experiment (positive-sum-only; 22 groups of 133 students; two-blocks of 50 rounds each paying either 2¢ or 20¢ per point) we modified the base game to give it a solution, a unique Nash equilibrium in the form of an "outside option" strategy. The SI discusses details of these and a third robustness experiment.

Qualitatively, groups in these Runway experiments tend in initial rounds to cluster on single-digit integers above zero before self-organizing into clusters that climb from round to round through the positive integers (Figure 2, full data Figure S16). Players hop up the strategy space with a median rate of 3 (mode 2; distribution in Figure S10). They coordinate overwhelmingly on the natural numbers, with only 4.2% of choices being negative or non-integer. They also flock: the median choice distance in a round was 7 (mode 2) and the median rate distance was 6 (mode 1). On these key qualities, behavior in the Runway Game is consistent with sophisticated flocking.

The effect of more rounds of experience is in all Runway experiments significantly larger hops in later rounds and increased flocking (tighter clustering; $\chi^2_1=14$; $\chi^2_1=458$; $\chi^2_1=593$; all $p<0.001$; Tables S8–S10). Larger groups and higher pay levels are also associated with larger hops ($\chi^2_1=328$, $\chi^2_1=17.4$; both $p<0.001$; Table S8). Zero-sum versions of the game show no difference in the magnitude of hopping and decreased clustering in choices and rates ($\chi^2_1=0.08$, $p=0.77$; $\chi^2_1=91$; $\chi^2_1=58$; both $p<0.001$; Tables S8–S10).

Comparing the results of the two Runway Game experiments, we find that players were indifferent to the presence of a Nash equilibrium, selecting it intentionally in less than 1% of rounds (SI) and exhibiting sophisticated flocking in both versions of the Runway Game.

*Beauty Contest*

The Beauty Contest is known for eliciting sophisticated iterated reasoning behavior. Since players in the repeated Beauty Contest have multiple rounds to potentially adapt their reasoning depths toward each other, we hypothesized that this game also fosters a previously unsuspected flocking dynamic. We reanalyzed data from two classic studies of the repeated Beauty Contest[38,39] to examine whether groups in these experiments had formed flocks. The first study administered 4 rounds with 7 groups of 15–18 people, and the second posed 10 rounds with 55 groups of 3–7.

It is established that choices in the Beauty Contest become closer to each other as they approach zero: from choice 50, iterating to depths one and two yields choices (33 and 22) that differ by 11, while starting the same processes from choice 5 yields outcomes that differ by just 1.1. However, the choice and rate clustering we propose is over and above this spurious type. To measure clustering due specifically to endogenous flock formation, controlling for the more trivial mechanism, we normalized choices and rates in each round relative to the benchmark set by the previous round's winning choice (SI). We find evidence for flocking over both experiments (via a significant increase in both choice and rate clustering; Figure 3, top 2 rows; $\chi^2_1$=25.3; $\chi^2_1$=22.5; both *p*<0.001; Table S14).

*Cross-game evidence for emergent flocking*

Figure 3 compares six experiments on three games—three experiments reported here, and three reanalyses—to give evidence that flocks emerge in multiple classes of games. A linear model over all three games finds increased flocking—lower choice distance and rate distance—with time ($\chi^2_1$=85.8; $\chi^2_1$=99.9; both *p*<0.001; Table S15). Across experiments, we see that flocks emerge with or without high pay levels, with or without Nash equilibria, and also that they are robust to group size, subject pool, and both within- and between-session experience (SI).

*Cross-game evidence for emergent within-group norms*

In games that elicit strategic iterated reasoning, players usually perform only 0–3 iterations.[49] Researchers have long debated whether these limited depths are due to players' own cognitive limits, or instead to the limits they expect of others.[38,43,50-53] We offer a third possibility for when games are repeated: expressed depths of iterated reasoning are a social norm that emerges, for each group, as a result of groupmates' previous patterns of mutual adaptation to each other. For example, given two players who adapt their depth of iterated reasoning toward each other over multiple rounds, with one initially expressing a depth of 1 and the other a depth of 5, a few rounds of experience will find them both expressing depths of 3. This norm-based theory of reasoning depth, however subtle, makes clear predictions that we support: 1) exhibited depths of reasoning (median hop size) should show great variability across groups but remain similar within groups (Figure 4), 2) an individual's depth of reasoning iterations should predict the median depth of the rest of their group ($\chi^2_1$=16.8; $p$<0.001), 3) the variation in members' reasoning depths should decrease with time ($\chi^2_1$=171; $p$<0.001), and 4) these patterns should disappear in randomly reshuffled groups (Table S15). Experiments that do not show evidence of this norm are the zero-sum Mod Game (perhaps because zero-sumness should suppress flocking[28]) and the 4-round Beauty Contest (which converges quickly). We conclude that each group develops its own customary limit on the number of strategic iterations that members expect each other to perform. This implicit social norm on cognitive sophistication develops in parallel with the sophisticated flocking collective behavior.

## Discussion

### *One flocking behavior in all games*

Players in the Mod Game, Runway Game, and Beauty Contest self-organize a common "sophisticated flocking" behavior, even though existing theory does not predict any similarity between these three multiplayer games. Since each game affords different formal solution concepts and is expected to evoke different kinds of reasoning, there is no reason to suspect any similarity in how they are played.

This argument is clearest for the games in their single-shot forms, but it is true of their repeated versions as well. While there is nothing in contemporary game theory that forbids one dynamic from appearing across many repeated games (well-known "folk theorems" permit virtually any pattern of behavior when certain games are repeated indefinitely[54]), failing to forbid a result is different from predicting it, much less explaining it. In this sense, established theory cannot account for our findings (SI).

### *Hopping due to iterated reasoning*

The most likely explanation for hopping is that players come to conceive of these games as evoking sophisticated oneupmanship via iterated best response. There are four reasons to support iterated reasoning as the explanation for hopping: 1) the most common rates in all three games were consistent with the 1–3 steps of iterated reasoning common in the literature;[49] 2) iterated reasoning provides a more parsimonious and psychological explanation than alternative theories (SI); 3) our findings in the Mod and Runway Games inspired our successful prediction of flocking in the Beauty Contest, in which iterated reasoning is already well-established as the cause of hopping; and 4) informally, participants often mentioned after sessions that they had used iterated reasoning to inform their choices.

Despite these arguments in favor of iterated reasoning in the Mod and Runway games, other evidence challenges them as a straightforward mechanism for hopping: 1) there is little theoretical justification for invoking iterated reasoning in both games, since neither game is amenable to the analytic procedure upon which the psychology of iterated reasoning is formally grounded ("iterated deletion of dominated strategies"), 2) players in some groups exhibit unrealistically high levels of iterated reasoning (up to 10 levels), and 3) despite the supposedly complex deliberative iterated reasoning process required for increasingly high rates, participants respond in the same unrealistically short amounts of time (within 10 seconds), regardless of how many iterations they performed.

These conflicts may be resolved if we introduce a distinction between iterated reasoning as a cognitive process and iterated reasoning as a "mental model." Suppose that a simple decision heuristic (such as "increment previous choice by *n*") can produce an "unsophisticated" hopping behavior superficially consistent with fully deliberative iterated "what you think I think you think I think" reasoning. If all players think of each other as applying the full process, but each in fact uses the cognitively cheap heuristic, then iterated reasoning is no longer applied by anyone as a cognitive procedure for generating behavior, but it continues to be applied by everyone as a way of understanding how others play the game.

### *An emergent mental model*

We take the reliable cross-game emergence of sophisticated flocking as evidence that the Mod Game, Runway Game, and Beauty Contest are all somehow eliciting the same mental model from participants.

A mental model is the conceptualization of a scenario within which one decides how to reason about it.[55,56] Human mental models operate at a very general level of reasoning, and, despite their importance, have had little concrete place in game theory. Their simplest expression in game theory is in their influence on whether a social decision is perceived as a game at all.[57,58] The mental models that govern our daily interactions seem to make strategic reasoning rare, even in situations that are formally strategic. For example, when you sit at a 4-way intersection, are you following the rules of a mundane non-strategic traffic procedure, or the dictates of a Machiavellian battle of wills? Each alternative conceptualization implies very different prescriptions for how a driver should reason and behave. Mental models can also make mundane decisions strategic merely by fearing that others are doing the same: surrounding a community of Nebraskan drivers with New Yorkers may make them less heedful, even where the rules of the road remain the same. Such self-fulfilling divergences between the game we see and the game we are in, between our mental and economic models, are ubiquitous, and certain to contribute to the complexity of social systems.

In the case of the three games we examine, a weaker, "mental-model-free" claim might suffice to explain emergent flocking in any game considered on its own, but the fact that all three engender the same very specific collective behavior seems to imply some unexpected conceptual similarity between the games, via a conceptual level of game understanding that intercedes between game theory and game behavior.

Whatever commonality the three games are ultimately found to share, it will not be found in their single-round Nash solutions. The mental models we observe seem to emerge without regard for the one-shot formal structures of these games. In the extreme case of the Runway Game, participants self-organize the same behavior regardless of whether or not the game even has a formal solution. It may be that large games, those with large numbers of players, choices, and rounds, can be harder to reason about even when they are as analytically tractable as simple games. Whatever the explanation, the insensitivity of players to formal structure challenges the relevance of elementary game-theoretic concepts to more complex games.

To explain the emergence of unexpected mental models in the Runway and Mod games, we hypothesize that noisy choices in early rounds incidentally create small clusters of choices —ephemeral or dynamically updating focal points—that break the strategic and perceptual symmetries of both games in a way that suggests sophisticated flocking. This feedback between observed behavior and mental model formation might explain our evidence that players' strategic iterations are limited by a social norm for how deeply one 'should' reason through the reasoning of others (Figure 4). One implication is that people may not be able to develop mental models of complex settings before actually experiencing them. If this is true, a general understanding of human game behavior will have to include mental models, and the processes by which they co-evolve with experience.

**Conclusion**

In repeated versions of three types of economic game, laboratory groups produce sophisticated flocking. Our tests demonstrate the robustness of this behavior over high and low payoffs, over multiple populations, and with very experienced players. These three game environments lack any formal resemblance to each other, so the emergence of one collective behavior in all three allows us to observe how mental models of social systems develop, establishing the importance of cognitive science and complex systems in economic behavior. We propose that mental model formation is an active process that can occur during game play. And because of the evidence that emergent social processes limit iterated reasoning (rather than the other way around), these games show that sophisticated human reasoning processes may be just as likely to drive the complex, oftentimes pathological, social phenomena that we usually attribute to reactive, emotional, non-deliberative reasoning. In other words, human intelligence may as likely increase as decrease the complexity and unpredictability of social and economic outcomes.

COGNITIVE MECHANISMS FOR FLOCKING 15

**Acknowledgements:**

The data are available upon request. The authors would like to thank Yuzuru Sato, Harmen De Weerd, Tatsuya Kameda, Arlington Williams, and James Walker. This work was supported in part by NSF REESE Grant 0910218; NSF/IGERT 0903495; NASA/INSGC Space Grant NNX10AK66H. This work is based on the dissertation of author SF (http://

gradworks.umi.com/35/99/3599175.html). All experiments were approved by the Indiana University IRB.

**Figures**

**Fig. 1:**

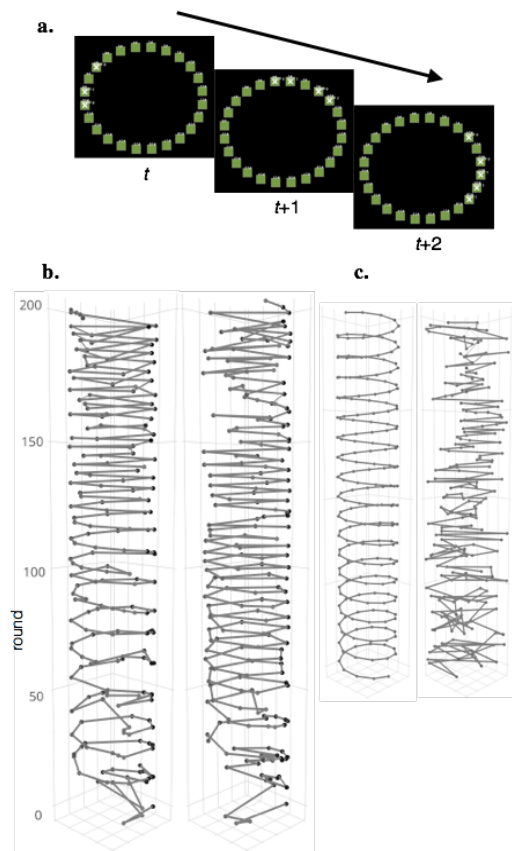

**Figure 1. Cycles in the Mod Game driven by sophisticated flocking.**

Panel **a.** shows three rounds of a five-player group in a cyclic flocking pattern. These players are flocking in the sense that they are proceeding together through the same sequence of choices over time. They are sophisticated in the sense that the flock does not advance incrementally through choices, but in larger hops. While reactive players best-respond to current state and advance through choices one at a time, sophisticated players best-respond to the anticipated best-responses of others, and advance more quickly in hops, as we observe. Panel **b.** plots 200 rounds of choices 1–24, laid out on a circular strategy space, for players from two groups. Panel **c.** depicts simulated data for comparison: a perfect flock with 200 hops of size 2 and a sequence of 200 randomly-generated integers 1–24. The spiral patterns in **b.** are consistent with the sophisticated flocking illustrated in the first plot of **c**. Figure S5 plots data from all groups (the data in **b.** are from groups 1 and 5).

**Fig. 2:**

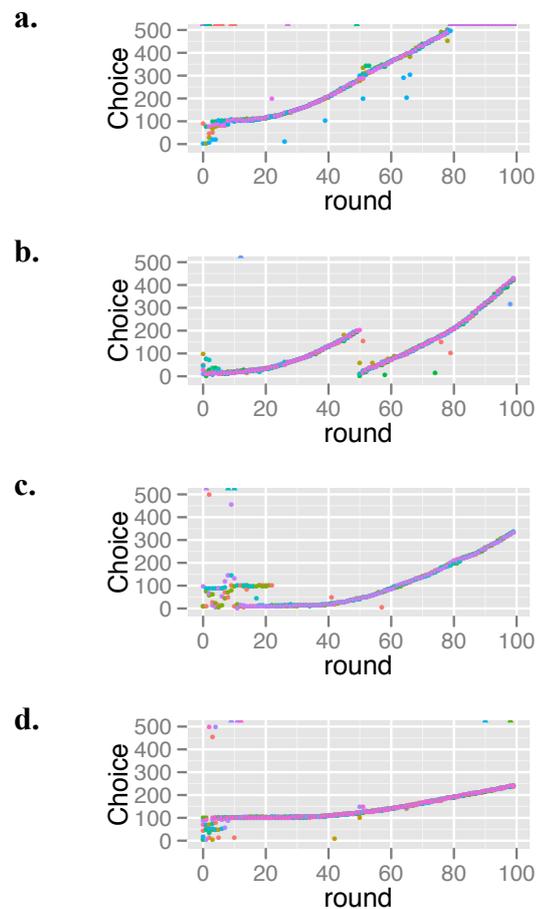

**Figure 2. Sophisticated flocking in four sessions of the Runway game.**

Plots show raw choices over 80 rounds. Each color of point represents a different group member. The apparent proliferation of points in initial rounds reflects disorganized behavior before groups have formed flocks, which appear in each plot as a line. The slopes of all lines are greater than 1, which can occur only when players are best-responding to the recursively anticipated best-responses of others. Furthermore, the different slopes of each panel suggest differing norms for how players come to expect each other to reason (Figure 4). These sessions used a two-block design, with each block lasting 40 rounds (this explains the "reset" in panel **b)**. Panels **a.–d.** represent groups 70–73 of Figure S17. Choices above 500 are truncated.

**Fig. 3:**

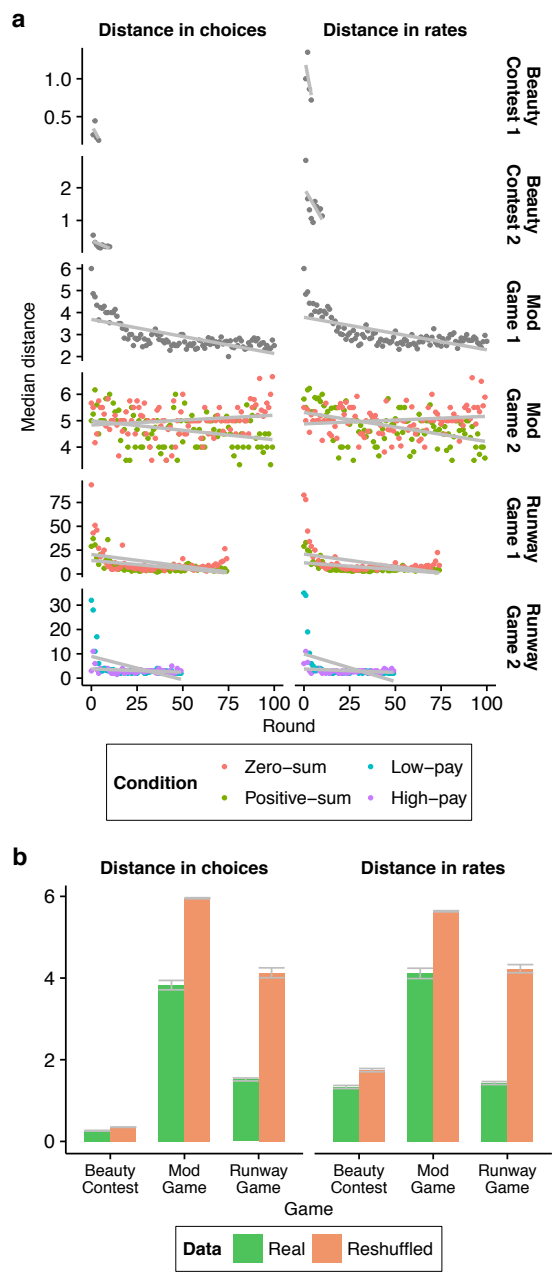

**Figure 3. Cross-game evidence for flocking.** We define flocking as within-group, within-round convergence on both choices (first column of both panels) and rates (or hops; second column). In the scatterplots of panel **a.**, each point represents a distance for a group in a round. These plots show that the distances between group members' choices and rates decrease over time, particularly in the initial rounds of play. Beauty Contests are limited in how many times they can be repeated, but flocking is clear in even these short time series. The first three rows of **a** (in grey) illustrate the flocking we identify in previously published data [24,38,39], and the last three plots (in color) show the data reported herein. Flocking increases with time (negative slopes) except in zero-sum conditions of the Mod and Runway games. Panel **b.** shows extent of flocking by comparing observed flocking to that in synthetic reshuffled datasets that randomly reassigned participants across groups. Bars show standard errors.

**Fig. 4:**

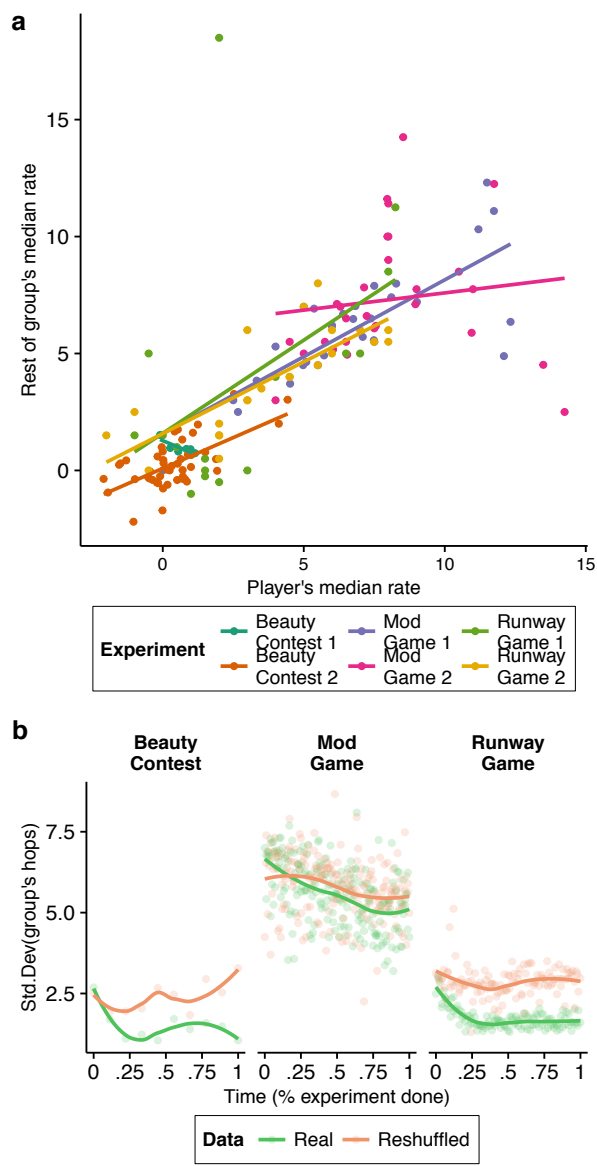

**Figure 4. Emergent group-level norms.**

Both panels present evidence that groups developed varying norms for the shared depth of recursive reasoning that their members exhibited. Greater hop sizes correspond to more deeply recursive strategic reasoning. Panel **a.** shows that median hop sizes exhibit both within-group homogeneity (slopes are ~1) and cross-group heterogeneity (axes have large domain and range). Panel **b.** shows that the variances in groups' hops decrease (green lines), particularly in the initial rounds of each game, and remains below that of groups that have been randomly reshuffled (orange).

**List of Supplementary information:**

— Materials and Methods
  — Measuring "Rate" and "hopping"
  — Mod Game SI
  — Runway Game SI
  — Beauty Contest SI
— Supplementary Text
  — Folk theorems offer an underspecified theory of game dynamics
  — Mental models
  — Norm formation
— Supplementary Figures
— Supplementary Tables
— Supplement References